\documentclass[12pt]{article}
\usepackage{graphicx}
\usepackage{amsmath}

\newcommand{\rcorr}{\hbox{\kern-1.2em$\longrightarrow$}}
\newcommand{\lrcorr}{\hbox{\kern-1.2em$\longleftrightarrow$}}

\newcommand{\nRightarrow}{\Rightarrow\kern-1.2em\hbox{/}\kern.8em} %
\newcommand{\BB}{\hbox{I\kern-.2em\hbox{B}}} 
\newcommand{\DD}{\hbox{I\kern-.2em\hbox{D}}} 
\newcommand{\FF}{\hbox{I\kern-.2em\hbox{F}}} 
\newcommand{\NN}{\hbox{I\kern-.2em\hbox{N}}}  
\newcommand{\ZZ}{{{\rm Z}\kern-.28em{\rm Z}}} 
\newcommand{\RR}{\mathop{{\rm I}\kern-.2em{\rm R}}\nolimits} 
\newcommand{\QQ}{\hbox{l\kern-.36em\hbox{Q}}}  
\newcommand{\CC}{\hbox{I\kern-.58em\hbox{C}}}

\begin{document}
\title{Measurements of Observables Replaced by ``Detections'' in Quantum Theory}
\author{Giuseppe Nistic\`o and Angela Sestito\footnote{Angela Sestito's work is supported by the European Commission,
European Social Fund and  by the Calabria Region, Regional Operative
Program (ROP) Calabria ESF 2007/2013 - IV Axis Human Capital -
Operative Objective M2- Action D.5}
\\
{\small Dipartimento
di Matematica, Universit\`a della Calabria, Italy}\\
{\small and}\\
{\small
INFN -- gruppo collegato di Cosenza, Italy}\\
{\small email: gnistico@unical.it} } \maketitle
\abstract{In Quantum Physics it is not always possible to directly perform the measurement of an obsevable;
in some of these cases, however, its value can be {\sl detected}, i.e. it can be inferred by measuring {\sl another} observable
characterized by perfect correlation with the observable of interest.
Though a {\sl detection} is often interpreted as a {\sl measurement} of the detected observable, we prove that the two concepts cannot be
identified in Quantum Physics.
Furthermore, we establish what meaning and role can be ascribed to detections coherently with Quantum Theory.
}
\section{Introduction}
In Quantum Physics there are circumstances where the measurement of some observables
encounters serious difficulties.
\par
In some cases the obstacle is the quantum incompatibility which forbids simultaneous measurements of
observables represented by non-commuting operators. This problem arises, for instance,
in a typical double-slit experiment; the position $Q_F$
of the particle on the final screen is incompatible with {\sl which slit} observable $E_S$
which takes value 1 if and only if the particle passes through the first slit; indeed,
the non-equality $[\hat E_S,\hat Q_F]\neq{\bf 0}$ holds for the operators
which represent $E_S$ and $Q_F$ \cite{1}.
\par
Another kind of obstacle can be caused by the particular features of the system under investigation.
For instance, if the system is a $\tau$ {\sl neutrino}, its very low interactivity
makes in practice impossible a direct measurement of its localization.
Physicists often outflank these obstacles by exploiting {\sl quantum correlations}.
A $\tau$ neutrino, under suitable conditions, provokes processes which yield
other particles, as $\tau$ particles, whose localization can be measured. So, under those conditions,
the localization of the $\tau$ particle is, ideally, perfectly correlated to the localization of the
$\tau$ neutrino in that region.
Then,
the localization (not measured) of the $\tau$ neutrino is {\sl inferred}
from an actually measured localization of the $\tau$ particle.
\par
The practice of making use of correlations to circumvent difficulties in measuring observables
can work also when the obstacle is the incompatibility between the observables of interest, as in the
double slit experiment where $[\hat E_S,\hat Q_F]\neq{\bf 0}$ forbids the knowledge of which slit
a particle localized on the final screen passed through. It has been
shown \cite{1} that, under suitable conditions, an observable $T_S$ exists such that $[\hat T_S,\hat Q_F]={\bf 0}$,
and whose outcomes are perfectly correlated
with the outcomes of $E_S$ in every simultaneous measurement of $T_S$ and $E_S$; therefore,
by measuring $T_S$ and $Q_F$ together, the position of the particle on the final screen is directly measured,
and which slit information is inferred from the outcome of $T_S$, exploiting the perfect correlation
between $T_S$ and $E_S$.
\par
Now, it is well known that serious inconsistencies can be derived in Quantum Mechanics by assigning
non measured observables values just by making use of correlations; valuable examples are the contradictions derived by
Greenberger, Horne, Shimony, Zeilinger (GHSZ) in \cite{2} and the pioneer theorem of Bell
\cite{3}.
Therefore, though these kind of outflanking methods are diffusely practised in Physics, the question
of establishing to what extent they can be interpreted as real measurements is not of secondary importance.
To overlook this problem submits the results of the scientific practice to risks of inconsistency.
\par
To distinguish {\sl assigning an observable $E$ the value obtained as actual outcome of a correlated
observable $T$} from a {\sl real measurement of $E$}, we call it {\sl detection} of $E$ by $T$.
The first question addressed in the present work is to investigate whether in Quantum Physics
the identification between detections and measurements is legitimate.
\par
To do this, in section 2 we
first develop the formalism related to the concepts of {\sl measurement} and {\sl detection},
within the standard interpretation of Quantum Mechanics.
Such a formalization makes possible to formulate the hypothetical identification of
a detection with the measurement
of the detected observable in terms of mathematical statements
within the theory.
\par
The theoretical consequences of the identification are investigated in section 3.
It soon appears clear that the identification affects the equivalence between co-measurability
of two observables and the commutativity of the corresponding operators.
In section 3.2 we prove that
assuming the identification is inconsistent with Quantum Theory, independently of the equivalence
of co-measurability with commutativity.
\par
This no-go achievement imposes a choice between two alternative options: either to insist in maintaining the identification
or to reject it. The first option would make necessary a new theory which replaces Quantum Theory.
\par
If the second option is followed, then the meaning and the role of detections, being not identifiable with measurements,
must be established coherently with Quantum Theory. Our work follows this last option.
To make our results independent of the
debate about the {\sl ontic nature} of Quantum Mechanics \cite{4},\cite{5}
we assume Quantum Theory as resulting
from the epistemic development
\cite{6},\cite{7},
\cite{8},\cite{9} of von Neumann's foundational approach \cite{10}.
\par
Our first step of section 4 easily shows that detections behave as {\sl perfect simulations} of the
measurement of the detected observable, in the sense that the physical consequences of
the outcomes of a measurement of the observable are indistinguishable from the consequences
of the outcomes of the detection of that observable.
\par
However, the interpretation of detections as simulations does not apply if the detection
is performed together with the measurement of an observable $F$ which is not compatible
with the detected observable; indeed, a measurement of the detected observable
could not even happen, and then there would be nothing to simulate.
In this more general case we prove that a language about physical events can be established,
which includes the events corresponding to the occurrences of the outcomes of the detectable,
but not measurable, observable. Detections supply a value assignment for the events in such a
language which is {\sl consistent}; i.e., the outcomes, the statistics and the correlations predicted
by this language are empirically and theoretically valid.
\section{Detections in Quantum Theory}
In section 2.3 we introduce the concept of detection within the quantum theoretical formalism.
To do this we need some
basic concepts of Quantum Theory which are explained in subsections 2.1 and 2.2.
\subsection{Basic Formalism}
Let $\mathcal H$ be the Hilbert space where the quantum theory of the physical system under investigation is formulated.
If $A$ is any observable, then by $\hat A$ we denote the self-adjoint operator which represents $A$; according to Quantum Theory,
the expected value of $A$ is $Tr(\rho\hat A)$, where
the density operator $\rho$ is the quantum state of the system.
\par
Given a quantum state $\rho$, by {\it support} of $\rho$ we mean any {\it concrete} subset ${\mathcal S}(\rho)$
of specimens of the
physical system \cite{11}, whose quantum state is represented by $\rho$.
Given a support
${\mathcal S}(\rho)$, by $\bf A({\mathcal S}(\rho))$ we denote the concrete subset of
all specimens in ${\mathcal S}(\rho)$ which {\sl actually}
undergo a measurement of $A$. In the following, we shall write simply $\bf A$ instead of
$\bf A({\mathcal S}(\rho))$ to avoid a cumbersome notation, whenever no confusion is likely.
\par
Let $E$ be any {\sl elementary} observables, i.e. an observable having
only two possible values denoted by $0$ and $1$, and hence
represented by a projection operator $\hat E$;
in this case the expected value $Tr(\rho\hat E)$ of $E$ coincides with the
probability that outcome $1$ occurs in a measurement of $E$.
The set of all elementary observables will be denoted by $\mathcal E$, while
$\hat{\mathcal E}({\mathcal H})$ denotes the set of all projection operators of
$\mathcal H$.
\par
Fixed any support ${\mathcal S}(\rho)$, in correspondence with every elementary
observable $E$ we define the following extensions of $E$ in ${\mathcal S}(\rho)$.
\begin{description}
\item[-]
the set ${\bf E}$ of the
specimens in ${\mathcal S}(\rho)$ which {\it actually} undergo a
measurement of $E$;
\item[-]
the set ${\bf E}_1$ of the specimens in ${\bf E}$ for which the
outcome $1$ of $E$ has been obtained;\par
\item[-]
the set ${\bf E}_0$ of the specimens in ${\bf E}$ for which the
outcome of $E$ is $0$.
\end{description}
To agree with Quantum Mechanics,
on the basis of the meaning of these concepts we have to assume that
the following statements hold (see \cite{11}, p.1268).
\begin{description}
\item[($2.1.i$)]
If $E$ is an elementary observable, then
for every $\rho$ a support
${\mathcal S}(\rho)$
exists  such that ${\bf E}\neq\emptyset$.
\item[($2.1.ii$)]
For every support ${\mathcal S}(\rho)$,
${\bf E}_1\cap{\bf E}_0=\emptyset$ and
${\bf E}_1\cup{\bf E}_0={\bf E}$, for every $\rho$.
\item[($2.1.iii$)]
If $Tr(\rho \hat E)\neq 0$ then
a support ${\mathcal
S}(\rho)$ exists such that ${\bf E}_1\neq\emptyset$,
and
\item[\qquad\;]
if $Tr(\rho \hat E)\neq 1$, then a support ${\mathcal
S}(\rho)$ exists such that ${\bf E}_0\neq\emptyset$.
\end{description}
\vskip.4pc
According to Standard Quantum Theory,
if the functional relation $\hat B=f(\hat A)$ holds between two
self-adjoint operators
$\hat A$ and $\hat B$ for a given real function $f$, then a measurement
of the observable $B$, henceforth denoted by $f(A)$, can be performed
by first measuring $A$ and then transforming the outcome $a$ by the function $f$ into the outcome $b=f(a)$ of $B$.
As a consequence, the following statement holds.
$$
\hbox{If}\quad B=f(A)\quad\hbox{then}\quad x\in{\bf A}\hbox{ implies }x\in{\bf B}.\eqno{(2.2)}
$$
If $[\hat A,\hat B]={\bf 0}$,
then a third self-adjoint operator $\hat C$ and two functions $f$ and $g$ exist so that
$\hat A=f(\hat C)$ and $\hat B=g(\hat C)$ \cite{10}. Thus
$A$ and $B$ can be measured together if the
corresponding operators commute with each other.
These implications and their consequences can be expressed in terms of extensions:
\begin{description}
\item[($2.1.iv$)] $\forall E,F\in{\mathcal E}$,\quad
$
[\hat E, \hat F]={\bf 0}\quad\hbox{ implies }
\quad\forall\rho\;\;\exists{\mathcal S}(\rho)\hbox{
such that }{\bf E}\cap{\bf F}\neq\emptyset$.
\item[($2.1.v$)]
If $[\hat A,\hat B]={\bf 0}$ and $\hat D=f(\hat A,\hat B)$
then $x\in{\bf A}\cap{\bf B}$ implies $x\in{\bf D}$, $\forall {\mathcal S}(\rho)$.
\item[($2.1.vi$)]
If $F$ and $G$ belong to $\mathcal E$ and $\hat F\hat G={\bf 0}$, then
${\bf F}_1\cap{\bf G}_1=\emptyset$, $\forall\rho$, $\forall
{\mathcal S}(\rho)$.
\end{description}
This last implication (2.1.vi) means that in every simultaneous measurement of $F$ and $G$ the outcome $1$ for $F$ and the outcome $1$ for $G$ are mutually
exclusive; in this case the projection $\hat E =\hat F+\hat G$ belongs to $\hat{\mathcal E}({\mathcal H})$; conversely,
if $\hat F+\hat G\in\hat{\mathcal E}({\mathcal H})$ for a pair $\hat F,\hat G\in\hat{\mathcal E}({\mathcal H})$, then
$\hat F\hat G={\bf 0}$. Hence, $\hat E =\hat F+\hat G$ represents an elementary observable $E$;
because of (2.1.iv), (2.1.v) a measurement's outcome of $E$ can be obtained by summing the outcomes of a simultaneous measurement of $F$ and $G$.
\subsection{Equivalence between co-measurability and commutativity}
Statement (2.1.iv) affirms that sufficient condition for the co-measurability of two
observables is the commutativity of the corresponding operators.
\par
The necessity of the commutativity between $\hat E$ and $\hat F$ for the
co-measurability of $E$ and $F$ is expressed by the following statement.
\begin{description}
\item[($2.3$)]
$
[\hat E, \hat F]\neq{\bf 0}\quad\hbox{implies}\quad {\bf
E}\cap{\bf F}=\emptyset\hbox{ for all }{\mathcal S}(\rho)
$.
\end{description}
Some treatises of Quantum Mechanics argue that if the state of the system is a common eigenstate of two self-adjoint operators
$\hat A$ and $\hat B$, i.e. if $\hat A\rho=a\rho$ and $\hat B\rho=b\rho$,
then the observables $A$ and $B$ represented by them are measurable together though $[\hat E,\hat F]\neq{\bf 0}$.
As valuable example we can cite von Neumann's book (\cite{10}, p. 230); accordingly, (2.3) would be violated.
\par
But the arguments supporting such a dependence of co-measurability on the state are based on the {\sl ontic} interpretation of the quantum state; in particular,
it is assumed that if a measurement of an observable $A$ yields the eigenvalue $a$ of $A$ as outcome, then the ({\sl ontic})
state of the system collapses into an eigenstate of $\hat A$ belonging to the eigenvalue $a$.
However, such an ontic interpretation is not necessary to a consistent interpretation of Quantum Theory
fully coherent with empirical observations.
In fact, conceptually coherent and well grounded axiomatic foundations of Quantum Mechanics \cite{6},\cite{7},
have been developed, where (2.3) is valid. Furthermore, (2.3) is decisive in shielding Quantum Mechanics \cite{11},\cite{12} from
inconsistencies with the locality principle, envisaged by the so called non-locality theorems \cite{2},\cite{3},\cite{13},\cite{14};
in other words, with (2.3) Quantum Mechanics keeps consistency also with locality.
\par
For these reasons, Standard Quantum Theory we refer to include the validity of (2.3).
\subsection{Detecting observables}
Experimental physicists must often resort to {\sl detect} observables, rather than {\sl directly}
measuring them. For instance,
one challenge of experimental particle physics is to test
the particle oscillation  from $\mu$ neutrinos to $\tau$ neutrinos \cite{15}-\cite{19}.
The OPERA experiment \cite{16,17} was
designed for testing that $\mu$ neutrinos created at CERN's laboratories in Geneva arrive, after the predicted oscillation, as $\tau$ neutrinos
at the laboratory LNGS of {\sl Gran Sasso} in Italy.
In order that the test be successful, it is necessary to localize the $\tau$ neutrinos at LNGS.
But neutrinos cannot be directly localized; in fact, in the experiment such a localization
is performed by localizing the $\tau$ particles emerging from a process
provoked by the $\tau$ neutrinos arriving from Geneva.
This {\sl correlation} between the two localizations, predicted by Quantum Mechanics,
is exploited to {\sl infer} the localization of the $\tau$ neutrino, not measured, from the actually
measured localization of the $\tau$ particle.
To formally distinguish such an inference from a direct measurement of the neutrino's localization, we call it a
{\sl detection}.
\par
Generalizing, given two elementary observables $E$ and $T$
we say that $E$ can be detected by $T$
if the perfect correlation {\sl the outcome of \,$T$ is $1$ iff the outcome of $E$ is $1$}
in every simultaneous measurement
holds according to Quantum Mechanics.
By {\sl detection} of $E$ we mean to assign $E$ the value obtained as outcome of
an actual measurement of $T$.
The existence of the correlation allowing for the detection depends on the quantum state of the system;
if such a correlation holds when the quantum state is $\rho$ we write $T\quad^\rho\lrcorr E$.
The existence of the Quantum Mechanical correlation $T\quad^\rho\lrcorr E$ implicitly implies
that $T$ and $E$ must be measurable together, of course.
\par
In the
OPERA experiment, the observable $T^{(\tau)}$ corresponding to the
localization of the $\tau$ particle is the observable actually measured, while
$E^{(\nu)}$ denotes the observable corresponding to the localization of the $\tau$ neutrino;
since the correlation $T^{(\tau)}\quad^\rho\lrcorr E^{(\nu)}$ holds, $E^{(\nu)}$ is detectable by $T^{(\tau)}$.
\vskip.5pc
The following general definition formalizes such a concept of detecting observable.
\vskip1pc\noindent
{\bf Definition 2.1.} {\sl The elementary observable $E$ is detectable by the elementary observable $T$ in the state $\rho$,
written $T\quad^\rho\lrcorr E$, if
\begin{description}
\item[{\rm (i)}] a support of $\rho$ exists such that ${\bf T}\cap {\bf E}\neq\emptyset$ (simultaneous measurability),
\item[{\rm (ii)}] for every specimen $x\in{\bf E}\cap{\bf T}$,
\item[]\qquad $x\in{\bf T}_1$ iff $x\in{\bf E}_1$\quad and \quad $x\in{\bf T}_0$ iff
$x\in{\bf E}_0$,
\item[\quad] holds in every support ${\mathcal S}(\rho)$.
\end{description}}
\noindent
In terms of probability,
the relation $T\quad^\rho\lrcorr E$ holds if and only if
in a simultaneous measurement of $T$ and $E$ both pairs of outcomes $(1,0)$ and $(0,1)$
have zero probability,
hence, in mathematical terms, if and only if $Tr(\rho\hat T[{\bf 1}-\hat E])={\bf 0}$ and $Tr(\rho[{\bf 1}-\hat T]\hat E)={\bf 0}$;
these two equations imply $\hat T\rho=\hat T\hat E\rho$ and $\hat E\rho=\hat E\hat T\rho$.
Therefore we have the following mathematical characterization of this relation.
\vskip1pc\noindent
{\bf Proposition 2.1.}
{\sl $T\quad^\rho\lrcorr E$ if and only if $[\hat T,\hat E]={\bf 0}$ and $\hat E\rho=\hat T\rho$.}
\section{Detections versus Measurements}
In the cited OPERA experiment, physicists interpret the presence of the $\tau$ particle in a point of LNGS
as a {\sl proof} that the $\tau$ neutrino which provokes the process is localized in a space-time neighborhood
of that point; i.e., if ${T^{(\tau)}}\quad^\rho\lrcorr {E^{(\nu)}}$, then
the measurement of $T^{(\tau)}$ {\sl is identified} with a measurement of $E^{(\nu)}$.
\par
If such an identification were valid, then
in general it should be possible to {\sl super-impose} the identification of
the outcome of a measurement of $T$ as outcome of $E$; then,
the following implication should hold in Quantum Theory.
$$
T\quad^\rho\lrcorr E\quad\hbox{implies}\quad {\bf T}_1\subseteq{\bf E}_1\hbox{ and }
{\bf T}_0\subseteq{\bf E}_0\,,\;\forall{\mathcal S}(\rho).\eqno(3.1)
$$
But the relation $T\quad^\rho\lrcorr E$ is symmetric; therefore,
the identification consists of the following statement.
$$
T\quad^\rho\lrcorr E\qquad\hbox{implies}\qquad {\bf T}_1={\bf E}_1\;\hbox{and}\;
{\bf T}_0={\bf E}_0\,,\;\forall{\mathcal S}(\rho).
\eqno(3.2)
$$
The aim of the present section is to test whether the identification (3.2) can be assumed to hold in Quantum Physics.
In the next subsection we show that Quantum Theory is not indifferent to the introduction of identification (3.2);
to assume (3.2) would entail an important revision of Quantum Theory.
But in section 3.2 we shall show that not even such a revision would make possible identification (3.2) in Quantum Physics.
\subsection{Theoretical Impact of the identification}
Let us explain how
the identification expressed by (3.2) would lead to a revision of the theoretical comprehension
of Quantum Mechanics.
\par
Let $E$ and $F$ be two elementary observables such that $[\hat E,\hat F]\neq{\bf 0}$.
Let $\hat C(\hat E,\hat F)$
be their {\sl commutation} projection, i.e. the projection operator which projects onto
the subspace ${\mathcal C}(\hat E,\hat F)=\{\psi\in{\mathcal H}\mid [\hat E,\hat F]\psi=0\}$
spanned by the common eigenvectors of $\hat E$ and $\hat F$ \cite{20}.
\par
It is not difficult to find mathematical examples of non-commuting
projection operators $\hat E$ and $\hat F$ with non-trivial commutation projection:
${\bf 0}\neq\hat C(\hat E,\hat F)\neq{\bf 1}$.
A physically relevant example of this situation is that pointed out by
Reiter and Thirring \cite{21} who found, in the Hilbert space
${\mathcal L}_2({\bf R})=\{\psi:{\bf R}\to {\bf C}\mid \int\vert\psi(x)\vert^2dx<\infty\}$ of the Quantum Theory
for a one-dimensional non relativistic particle, a wave function $\psi_{RT}$ satisfying
$\hat E\psi_{RT}=\psi_{RT}$ and $\hat F\psi_{RT}=\psi_{RT}$, where $\hat E$ and $\hat F$ are non-trivial projection operators
respectively belonging to the spectral family of the position operator $\hat Q$ and to the spectral family
of the momentum operator $\hat P=-i\frac{\partial}{\partial x}$.
So we have $[\hat E,\hat F]\neq{\bf 0}$, but $\psi_{RT}\in{\mathcal C}(\hat E,\hat F)\neq\{\underline o\}$, so that
${\bf 0}\neq\hat C(\hat E,\hat F)\neq{\bf 1}$.
Now, the rank-one
projection $\hat T=\vert\psi_{_{RT}}\rangle\langle\psi_{_{RT}}\vert$ represents an elementary observable $T$;
Reither and Thirring proved that $\hat T\psi_{_{RT}}=\psi_{_{RT}}=\hat E\psi_{_{RT}}=\hat F\psi_{_{RT}}$ must hold. Therefore
$T\quad^\rho\lrcorr E$ holds, where $\rho=\vert\psi_{_{RT}}><\psi_{_{RT}}\vert$; thus,
${\bf T}={\bf E}={\bf F}$ for every ${\mathcal S}(\rho)$ would follow from (3.2).
But $[\hat E,\hat F]\neq{\bf 0}$ holds, being ${\bf E}\cap{\bf F}={\bf F}\neq\emptyset$, contrary to (2.3).
The violation of (2.3) can be obtained following this argument for every pair $E,F\in{\mathcal E}$ such that
$[\hat E,\hat E]\neq{\bf 0}$ and ${\bf 0}\neq\hat C(\hat E,\hat F)\neq{\bf 1}$.
\par
So, if the detection of an observable were identifiable with its measurement,
then Theoretical Physics would be forced to reconsider the
equivalence between co-measurabi\-lity of observables and commutativity of the corresponding operators,
by allowing some pairs of non-commuting observables for being measurable together in suitable states.
The ensuing revision of the interpretation of Quantum Theory would be not marginal.
In fact, a feature of valuable conceptually coherent and well grounded
axiomatic foundations of Quantum Mechanics, such as \cite{6},\cite{7} is that
the co-measurability relation is {\sl independent} of the quantum state; rather,
it is a property ascribable to the pair $(E,F)$ of the involved observables.
Furthermore, (2.3) is decisive for keeping the consistency of
Quantum Mechanics with locality \cite{11},\cite{12}.
\par
Thus, the validity of the identification (3.2) would make necessary a revision of the
conceptual bases of Quantum Theory and of its theoretical achievements.
Such a revision should consists in re-developing the theory starting from the removal of (2.3) and
the introduction of (3.2).
\subsection{Impossibility of the identification}
In this subsection we prove that it is not possible to re-develop Quantum Theory by removing (2.3) and introducing (3.2);
indeed we prove that introducing (3.2) leads to contradictions {\sl independently of the validity of (2.3)};
thus we have to conclude that the envisaged revision is impossible
in Quantum Physics\footnote{Our argument makes use of the mathematical setting adopted by GHSZ \cite{2} to prove that a given
set of conceptual hypotheses (local hidden variables) are in contradiction with Quantum Theory.
In \cite{11} we proved that if the hypotheses of GHSZ
are modified, then the contradiction does not necessarily arise. The hypotheses of the present argument, i.e.
identification (3.2), are different from those of GHSZ. Thus the occurrence of a contradiction needs an explicit proof.}.
\vskip.5pc
Let $a^\alpha,a^\beta,b,c^\alpha,c^\beta,d^\alpha,d^\beta$ be seven real numbers.
Greenberger, Horne, Shimony and Zeilinger (GHSZ) \cite{2} proved that
the following constraints
$$
\left\{
\begin{array}{llll}
\textrm{   i)}\quad &{a}^\alpha{b}&=-{c}^\alpha{d}^\alpha,\\
\textrm{  ii)}\quad &{a}^\beta{b}&=-{c}^\beta{d}^\alpha,\\
\textrm{ iii)}\quad &{a}^\beta{b}&=-{c}^\alpha{d}^\beta,\\
\textrm{  iv)}\quad &{a}^\alpha{b}&={c}^\beta{d}^\beta.\\
\end{array}\right.
\eqno(3.3)
$$
are not consistent with the further constraint that each of the
seven numbers must be $+1$ or $-1$. Indeed, (3.3.i) and (3.3.iv) imply
$-c^\alpha d^\alpha=c^\beta d^\beta$, while (3.3.ii) and (3.3.iii) imply
$-c^\beta d^\alpha=-c^\alpha d^\beta$; under the further constraint,
the product of these two equations is $c^\alpha c^\beta=-c^\alpha c^\beta$,
which is contradictory.
\par
Now we shall single out seven observables
$A^\alpha,A^\beta,B,C^\alpha,C^\beta,D^\alpha,D^\beta$
of a particular quantum system, whose possible outcomes can be
$+1$ or $-1$.
They are chosen so that they are all measurable together {\sl if the identification (3.2) holds}.
Then we show that their simultaneous outcomes $a^\alpha,a^\beta,b,c^\alpha,c^\beta,d^\alpha,d^\beta$
must satisfy the constraints (3.3) which are not consistent. Therefore we must conclude
that the identification of measurements with detection is not possible in Quantum
Mechanics.
\vskip.5pc
To realize such a program,
we consider a quantum system described in the Hilbert space
${\mathcal H}={\mathcal H}_1\otimes{\mathcal H}_2\otimes{\mathcal H}_3\otimes{\mathcal H}_4$,
where each ${\mathcal H}_k$ is ${\bf C}^2$. Then seven elementary observables are defined by
the following projection operators which represent them.
\par
$\hat E^\alpha=\frac{1}{2}\left[\begin{array}{cc}
1 & 1 \\
1 & 1 \\
\end{array}\right]_1\otimes{\bf 1}_2\otimes{\bf 1}_3\otimes{\bf 1}_4$;\qquad
$\hat E^\beta=\frac{1}{2}\left[\begin{array}{cc}
1 & -i \\
i & 1 \\
\end{array}\right]_1\otimes{\bf 1}_2\otimes{\bf 1}_3\otimes{\bf 1}_4$;
\par
$\hat F={\bf 1}_1\otimes\frac{1}{2}\left[\begin{array}{cc}
1 & 1 \\
1 & 1 \\
\end{array}\right]_2\otimes{\bf 1}_3\otimes{\bf 1}_4$;\par
$\hat G^\alpha={\bf 1}_1\otimes{\bf 1}_2\otimes\frac{1}{2}\left[\begin{array}{cc}
1 & 1 \\
1 & 1 \\
\end{array}\right]_3\otimes{\bf 1}_4$;\qquad
$\hat G^\beta={\bf 1}_1\otimes{\bf 1}_2\otimes\frac{1}{2}\left[\begin{array}{cc}
1 & -i \\
i & 1 \\
\end{array}\right]_3\otimes{\bf 1}_4$;\par
$\hat L^\alpha={\bf 1}_1\otimes{\bf 1}_2\otimes{\bf 1}_3\otimes\frac{1}{2}\left[\begin{array}{cc}
1 & 1 \\
1 & 1 \\
\end{array}\right]_4$;\qquad
$\hat L^\beta={\bf 1}_1\otimes{\bf 1}_2\otimes{\bf 1}_3\otimes\frac{1}{2}\left[\begin{array}{cc}
1 & -i \\
i & 1 \\
\end{array}\right]_4$;
\vskip.5pc\noindent
Let the physical system be assigned the pure state $\rho_0=\vert\psi_0\rangle\langle\psi_0\vert$,
where
$$
\psi_0=\frac{1}{\sqrt{2}}\left(\left[\begin{array}{c}
1  \\0
\end{array}\right]_1
\otimes\left[\begin{array}{c}
1  \\0
\end{array}\right]_2
\otimes\left[\begin{array}{c}
0  \\1
\end{array}\right]_3
\otimes\left[\begin{array}{c}
0  \\1
\end{array}\right]_4
-\left[\begin{array}{c}
0  \\1
\end{array}\right]_1
\otimes\left[\begin{array}{c}
0  \\1
\end{array}\right]_2
\otimes\left[\begin{array}{c}
1  \\0
\end{array}\right]_3
\otimes\left[\begin{array}{c}
1  \\0
\end{array}\right]_4
\right).
$$
Now we introduce the seven observables
$A^\alpha=2E^\alpha-1$, $A^\beta=2E^\beta-1$, $B=2F-1$,
$C^\alpha=2G^\alpha-1$, $C^\beta=2G^\beta-1$,
$D^\alpha=2L^\alpha-1$, $D^\beta=2L^\beta-1$.
The possible outcomes of each of these observables are $+1$ or $-1$.
\par
We show that {\sl if the identification (3.2) is assumed to hold}, then all
observables $A^\alpha,A^\beta,B,C^\alpha,C^\beta,D^\alpha,D^\beta$ are measurable together
on a specimen $x_0$ and that
their respective outcomes $a^\alpha,a^\beta,b,c^\alpha,c^\beta,d^\alpha,d^\beta$
should satisfy the {\sl contradictory} constraints (3.3).
\par
The four projection operators $\hat E^\alpha,\hat F,\hat G^\beta,\hat L^\alpha$ commute with
each other; hence, all the corresponding elementary observables can be measured together, i.e.
a support ${\mathcal S}(\rho_0)$ and a specimen $x_0\in{\mathcal S}(\rho_0)$ exist such that
$$
x_0\in {\bf E}^\alpha\cap{\bf F}\cap{\bf G}^\beta\cap{\bf L}^\alpha.
$$
Let $\eta^\alpha,\phi,\gamma^\beta,\lambda^\alpha$ be the respective
outcomes of the measurements of $E^\alpha,F,G^\beta,L^\alpha$ on such a specimen $x_0$.
Since
$A^\alpha=2E^\alpha-1$, $B=2F-1$,
$C^\beta=2G^\beta-1$,
$D^\alpha=2L^\alpha-1$, by (2.2) we deduce
$$x_0\in{\bf A}^\alpha\cap{\bf B}\cap{\bf C}^\beta\cap{\bf D}^\alpha
$$
and
$a^\alpha=2\eta^\alpha-1$, $b=2\phi-1$, $c^\beta=2\gamma^\beta-1$, $d^\alpha=2\lambda^\alpha-1$
must be the respective outcomes.
\par
Now, the projection operator
$$
\hat M=\frac{{\bf 1}-\hat A^\alpha\hat B\hat D^\alpha}{2}
$$
is a function of $\hat E^\alpha,\hat F,\hat L^\alpha$; therefore $x_0\in{\bf M}$ because of (2.1.v)
and $\mu=\frac{1}{2}(1-a^\alpha b d^\alpha)$ must be the outcome of the elementary observable $M$
measured on $x_0$.
\par
But
$[\hat M,\hat G^\alpha]={\bf 0}$ trivially holds; moreover, a direct calculation shows that
the equation $\hat M\rho_0=\hat G^\alpha\rho_0$ is satisfied; then the identification (3.2) implies
$x_0\in {\bf G}^\alpha$ and $\gamma^\alpha=\mu=\frac{1}{2}(1-a^\alpha b d^\alpha)$ is to be identified
as the outcome of the measurement of $G^\alpha$ on $x_0$. Since $C^\alpha=2G^\alpha-1$, by (2.1)
$x_0\in{\bf C}^\alpha$ holds too and
$c^\alpha=2\gamma^\alpha-1=-a^\alpha b d^\alpha$ is the outcome of $C^\alpha$; then
$x_0\in{\bf A}^\alpha\cap{\bf B}\cap{\bf C}^\alpha\cap{\bf C^\beta}\cap{\bf D^\alpha}$ and the
constraint
(3.3.i) must hold for the simultaneous measurement of $A^\alpha,B,C^\alpha,C^\beta,D^\alpha$
on the specimen $x_0$.
\par
Now we derive (3.3.ii).
By defining $\hat N=\frac{{\bf 1}-\hat B\hat C^\beta\hat D^\alpha}{2}$,
we can verify that  $[\hat N,\hat E^\beta]={\bf 0}$ and $\hat N\rho_0=\hat E^\beta\rho_0$ hold.
Then,
following the argument which led us to (3.3.i), with $N$ replacing $M$ and $A^\beta$ replacing
$C^\alpha$, we obtain that $x_0\in{\bf A}^\beta$ holds too and that
$a^\beta=-b c^\beta d^\alpha$ is the outcome of $A^\beta$ on $x_0$. Then
$x_0\in{\bf A}^\alpha\cap{\bf A}^\beta\cap{\bf B}\cap{\bf C}^\alpha\cap{\bf C^\beta}\cap{\bf D^\alpha}$ and the
constraint
(3.3.ii) must hold for the simultaneous measurement of $A^\alpha,A^\beta,B,C^\alpha,C^\beta,D^\alpha$
on the specimen $x_0$.
\par
Similarly, we derive (3.3.iii). Once defined
$\hat R=\frac{{\bf 1}-\hat A^\beta\hat B\hat C^\alpha}{2}$,
we can verify that  $[\hat R,\hat L^\beta]={\bf 0}$ and $\hat R\rho_0=\hat L^\beta\rho_0$ hold.
From $D^\beta=2L^\beta-1$ we obtain that $x_0\in{\bf D}^\beta$ holds too and that
$d^\beta=-a^\beta b c^\alpha$ is the outcome of $D^\beta$ on $x_0$. Then
$$x_0\in{\bf A}^\alpha\cap{\bf A}^\beta\cap{\bf B}\cap{\bf C}^\alpha\cap{\bf C^\beta}\cap{\bf D^\alpha}\cap{\bf D}^\beta$$
and the
constraint
(3.3.iii) must hold for the simultaneous measurement of $A^\alpha$, $A^\beta$, $B$,
$C^\alpha$, $C^\beta$, $D^\alpha$, $D^\beta$
on the specimen $x_0$.
\par
But we can also define
$\hat S=\frac{{\bf 1}+\hat A^\alpha \hat B\hat C^\beta}{2}$; therefore $x_0\in{\bf S}$, by (2.v),
and $\sigma=\frac{1}{2}(1+a^\alpha b c^\beta)$ must be the outcome of the elementary observable $S$
measured on $x_0$.
Now,
$[\hat S,\hat L^\beta]={\bf 0}$ trivially holds; moreover, the equation $\hat S\rho_0=\hat L^\beta\rho_0$
turns out to be satisfied; then the identification (3.2) implies that
$\lambda^\beta=\sigma=\frac{1}{2}(1+a^\alpha b c^\beta)$ is to be identified
as the outcome of $L^\beta$ measured on $x_0$. Since $D^\beta=2L^\beta-1$, by (2.1)
$d^\beta=2\lambda^\beta-1=a^\alpha b c^\beta$ is the outcome of $D^\beta$; then all the
constraints
(3.3) must hold for the simultaneous measurement of $A^\alpha,A^\beta,B,C^\alpha,C^\beta,D^\alpha,D^\beta$
on the specimen $x_0$.
\vskip1pc
Therefore, condition (3.2) which identifies a measurement of $E$ with its detection by a detecting
observable $T$ is not consistent with Quantum Mechanics.
Thus,
identification (3.2) cannot be super-imposed to Quantum Mechanics.
\section{Interpreting Detections}
In section 3
we attained the conclusion that in Quantum Physics detections of an observable $E$
by another observable $T$ cannot be identified with authentic measurements of $E$.
This conclusion
exempts Theoretical Physics from the revision of Quantum Mechanics envisaged in section 3.1; however,
since the Theory is kept unaltered, the task
of sharply identifying the tie between a detection by $T$ and the detected observable $E$
{\sl coherently with Quantum Theory} becomes unavoidable.
\par
The present section is our contribution towards an answer to this problem.
The first quick step of section 4.1 identifies a detection as a perfect {\sl measurement's simulation}
of the detected observable.
\par
However, if the detecting observable $T$ is measured together with an observable $F$
incompatible with the detected observable $E$, i.e. if $[\hat E,\hat F]\neq{\bf 0}$,
the interpretation as simulation cannot be maintained. In sections 4.2 and 4.3
we address the problem of establishing what role can be assigned detections in this more general case. We find
that detections have the role of providing Quantum Physics with a value assignment for $E$ beyond
the assignment provided by measurements,
which is consistent, in the sense that it introduces no
risk of contradictions, and can be used to draw conclusion
valid both from an empirical and form a theoretical point of view.
\subsection{Detections as `simulations' of measurements}
A way to understand how the detections of an elementary observable $E$
by $T$ and the measurement of $E$ are tied is to
compare the {\sl physical consequences} of the occurrences of the outcomes of $E$
with the physical consequences of the occurrences of the corresponding outcomes of $T$.
The physical consequences of the occurrence of an outcome of $T$ (resp., $E$)
are made explicit by the {\sl correlations} between the actually measured outcomes of {\sl another}
elementary observable $F$ and the occurrence of the outcomes of $T$ (resp., $E$).
Theoretically, these correlations are expressed by the concept of {\sl conditional probability}
in Quantum Mechanics.
\par
If $F$ and $G$ are elementary observables such that $[\hat F,\hat G]={\bf 0}$,
then the real number
$P(F\mid G)=Tr(\rho \hat F\hat G)/Tr(\rho \hat G)$ is the probability of occurrence of outcome $1$
for $F$ under the condition that outcome $1$ for $G$ occurs.
Analogously, if $G'$ denotes the elementary observable represented by the projection operator
${\hat G}'\equiv{\bf 1}-\hat G$, then
$P(F\mid G')=Tr(\rho \hat F{\hat G}')/Tr(\rho {\hat G}')$ is the probability of occurrence of outcome $1$
for $F$ under the condition that outcome $0$ for $G$ occurs
\par
Then, the comparison between the physical consequences of the occurrences of the outcome of $T$
with the physical consequences of the occurrences of the outcomes of $E$ can be carried out by comparing
the conditional probabilities $P(F\mid T)$, $P(F\mid T')$ with the conditional probabilities
$P(F\mid E)$, $P(F\mid E')$; these conditional probabilities are defined whenever
$[\hat F,\hat T]=[\hat F,\hat E]={\bf 0}$; therefore the domain of the comparison is
the following set of elementary observables
$$
{\mathcal F}_E(T)=\{F\in {\mathcal E}\mid  [\hat F,\hat T]=[\hat F,\hat E]={\bf 0}\}.
$$
Now, if $T\quad^\rho\lrcorr E$, then
the following statement follows from prop. 2.1.
$$
P(F\mid T)=\frac{Tr(\rho\hat F\hat T)}{Tr(\rho\hat T)}=\frac{Tr(\rho\hat F\hat E)}{Tr(\rho\hat E)}=
P(F\mid E), \quad \forall F\in{\mathcal F}_E(T).\eqno(4.1.i)
$$
Now, by making use of prop. 2.1 we easily obtain that
$T\quad^\rho\lrcorr E$ holds iff $T'\quad^\rho\lrcorr E'$; so (4.1.i) can be extended to
$$
P(F\mid T')=\frac{Tr(\rho\hat F{\hat T}')}{Tr(\rho{\hat T}')}=\frac{Tr(\rho\hat F{\hat E}')}{Tr(\rho{\hat E}')}=
P(F\mid E'), \quad \forall F\in{\mathcal F}_E(T).\eqno(4.1.ii)
$$
Then, if $T$ detects $E$, the effects of the occurrence of an outcome of $E$ are indistinguishable from the effects of the occurrence of
the same outcome of $T$.
In such a precise sense, we can conclude that the measurement of $E$ is perfectly {\sl simulated}
by a measurement of a detecting observable $T$.
\subsection{To Detect $E$ while measuring incompatible observables}
Let $T$, $E$ and $F$ be elementary observables
such that
$T\quad^\rho\lrcorr E$ and $[\hat T,\hat F]={\bf 0}$, so that $T$ can be measured together with $F$;
in the case in which $[\hat F,\hat E]\neq{\bf 0}$,
a measurement of $T$ simultaneous to a measurement of $F$
cannot be interpreted as a simulation of a measurement of $E$, according to section 4.1,
because
the possibility of a measurement of $E$ is forbidden in Quantum Physics, and hence there is nothing
to simulate.
Therefore, the problem arises of identifying the meaning to be ascribed to a detection
of $E$ by $T$ simultaneous to a measurement of an observable $F$ with $[\hat F,\hat E]\neq{\bf 0}$.
\vskip1pc\noindent
{\bf Example 4.1}
An emblematic circumstance of this situation is the two slit experiment. Let
$E_S$ be the {\sl which slit} elementary observable whose outcome is $1$ (resp., $0$) if the particle is
measured to be localized in the
first slit (resp., in the second slit) at the time in which it crosses the panel supporting the slits.
For each particle whose final localization in a given region $\Delta$
of the final screen is actually measured, represented by the projection operator $\hat F(\Delta)$,
it is impossible \cite{1} to measure the {\sl which slit} observable $E_S$ because $[\hat E_S,\hat F(\Delta)]\neq{\bf 0}$.
To outflank such an obstacle several methods were conceived over the years,
such as the recoiling slit, by Einstein \cite{22}, the light-electron
scattering scheme, by Feynman \cite{23}, the micro-maser apparatus, by Englert, Scully and
Walther (ESW) \cite{24}-\cite{26}.
Each of these methods can be theoretically described \cite{1} by introducing
an elementary observable $T$
{\sl which detects} $E$ according to prop. 2.1; the perfect correlation between the outcomes
of a simultaneous measurement of $T$ and $E$
allows to ascertain the outcome of $E$ ({\sl which slit})
by looking only at the outcome of $T$.
But $T$ is chosen so that $[\hat T, \hat F(\Delta)]={\bf 0}$ holds too;
the outflanking method prescribes of measuring $T$ and $F(\Delta)$
together, but not $E_S$; then, from the occurrence of outcome $1$ (resp., $0$) for $T$ it is inferred
that ``the particle passed through the first slit (resp., the second slit)''.
\vskip1pc\noindent
However, we saw in section 3.2 that the final inference in example 4.1 {\sl cannot} be interpreted as a
real measurement of $E_S$.
Such a kind of interpretative lack occurs in a general situation where
$$
\hat T\rho=\hat E\rho,\quad[\hat T,\hat F]={\bf 0}, \quad [\hat T,\hat E]={\bf 0}\quad\hbox{but}\quad
[\hat E,\hat F]\neq{\bf 0}.
\eqno(4.2)
$$
Hence our problem consists in identifying the role of
a detection of $E$ by $T$ simultaneous to a measurement of $F$, when $[\hat F,\hat E]\neq{\bf 0}$,
{\sl coherently with Quantum Theory}.
\vskip.5pc
Here we address such a problem.
Our achievements can be summarized as follows.
\par
First, we establish conditions which make
assigning $E$ {\sl a value} (not necessarily the outcome of $T$),
in a joint measurement of $T$ and of whatever observable $F$,
{\sl consistent} with the real outcomes' occurrences of all {\sl actually measured observables}.
\par
Then, we exploit a result of Cassinelli and Zangh\`i to conclude
that these conditions lead to a unique explicit probability ruling over such a
consistent value assignment jointly
with the occurrences  of
actual measurements' outcomes.
\par
Finally, we show that this unique probability coincides with that obtained from assigning $E$
just the value coinciding with the outcome of $T$; this particular assignment shall be called {\sl assignment by detection}.
\par
In general, the meaning of the value assignment by detection amounts to the fact that
it allows to assign events corresponding to the occurrence of the outcomes of
$E$ values simultaneous to the outcomes of whatever actually measured observable $F$.
The consistency proved by our achievements, whose exact meaning is made explicit in the following subsection 4.3,
implies that
all conclusions drawn from using this assignment are valid conclusions, both from
a theoretical and from an empirical point of view, and are contradictions free.
\par
Hence,
in the particular case of
the {\sl two slit} experiment of example 4.1, we can state that,
though the interpretation of outcome 1 (resp., 0) of observable $T$ detecting which slit
observable $E$ as the statement ``the particle passed through slit $1$ (resp., $0$)''
is not possible, because of the no-go proof of section 3.1, to give that statement validity
does not provoke contradictions, and all conclusions drawn from
using such a sentence are valid, either theoretically and empirically.
This explains the absence of interference in these experiments \cite{1},\cite{27}.
\subsection{Derivation of the results}
Let $T$ and $E$ be two elementary observables such that $T\quad^\rho\lrcorr E$.
To single out explicit conditions which make
assigning $E$ a value, in a joint measurement of $T$ and of another elementary observable $F$,
consistent with the real occurrences of measurements' outcomes,
it is worth to introduce a probability
$$
p_\rho(E\&\cdot):{\mathcal F}\to [0,1]\,,\; F\to p_\rho(E\&F).
$$
which should rule over the joint occurrence of
\begin{description}
\item[-]
outcome 1 from an actually performed measurement of $F$,
and
\item[-]
the assignment of value 1 to $E$.
\end{description}
To be coherent with Quantum Mechanics,
a first consistency condition for the probability $p_\rho(E\&\cdot)$ ruling over such a value assignment
is that, whenever it exists,
$p_\rho(E\&\cdot)$ must be defined on
the set ${\mathcal F}(T)=\{F\in {\mathcal E}\mid [\hat F,\hat T]={\bf 0}\}$.
i.e. ${\mathcal F}={\mathcal F}(T)$.
\par
A further consistency condition with real measurements' outcomes have to
establish that whenever $E$ could be actually measured,
i.e. if $[\hat E,\hat F]={\bf 0}$, then the probability $p_\rho(E\&\cdot)$ ruling over the value assignment must
coincide with the probability ruling over the occurences of the outcome of $E$ and $F$.
Since
the predictions of Quantum Mechanics about actually performed measurements are
empirically valid, once introduced
$\hat{\mathcal F}(\hat T)=\{\hat F\in \hat{\mathcal E}\mid [\hat F,\hat T]={\bf 0}\}$,
we require the following conditions.
\begin{description}
\item[{\rm (C.1.a)}]
if $ F\in{\mathcal F}_E(T)\subseteq{\mathcal F}(T)$, i.e. if $[\hat E,\hat F]={\bf 0}$,
then $p_\rho(E\& F)=Tr(\rho\hat E\hat F)$.
\item[{\rm (C.2.a)}]
if
$\{F_j\}_{j\in J}\subseteq {\mathcal F}(T)$ is any countable family such that
$\sum_j\hat F_j\equiv \hat F\in\hat{\mathcal F}(\hat T)$, then
\quad$p_\rho( E\& F)=\sum_{j\in J}p_\rho( E\& F_j)$.
\end{description}
Condition (C.1.a) requires that probability $p_\rho(E\& F)$ extends $Tr(\rho\hat E\hat F)$, i.e. the probability
of actual occurrence of outcome $1$ for both $E$ and $F$, with $F\in{\mathcal F}_F(T)$.
\par\noindent
Condition (C.2.a) can be inferred from the fact that, according to (2.1.vi), the outcome of $F$ can be obtained as the sum of
simultaneous outcomes of all $F_j$'s.
\vskip.5pc\noindent
Now, by making use of the following results \cite{28} of
Cassinelli and Zangh\`i,
we prove that such a probability there exists and it is unique.
\vskip.8pc\noindent
{\bf Theorem 4.1.} {\sl
Let $\hat{\mathcal A}$ be any von Neumann
algebra\footnote{A Von Neumann algebra \cite{29}
is a subset $\hat{\mathcal A}$ of bounded linear operators of the Hilbert space $\mathcal H$
such that $\hat{\mathcal A}=(\hat{\mathcal A}^\prime)^\prime\equiv\hat{\mathcal A}^{\prime\prime}$, where $\hat{\mathcal A}'$ denotes the
{\sl commutant} of $\hat{\mathcal A}$, i.e. the set of all bounded linear operators $\hat B$ of $\mathcal H$
such that $[\hat B,\hat A]={\bf 0}$ for all $\hat A\in\hat{\mathcal A}$.
The theory of Von Neumann algebras \cite{29} shows that if $\hat\Pi(\hat{\mathcal A})$ is the set
of all projection operators in the Von Neumann algebra $\hat{\mathcal A}$,
then $\hat{\mathcal A}=\hat\Pi(\hat{\mathcal A})^{\prime\prime}$.
}
of the Hilbert space $\mathcal H$, and let
$\Pi(\hat{\mathcal A})\subseteq\hat{\mathcal E}$ the set of all projection operators in $\hat{\mathcal A}$.
\par
If $\alpha:\Pi(\hat{\mathcal A})\to [0,1]$ is a normalized ($\alpha({\bf 1})=1$) functional
satisfying $\alpha(\sum_{j\in J}\hat F_j)=\sum_{j\in J}\alpha(\hat F_j)$ for every countable
family $\{\hat F_j\}_{j\in J}\subseteq\Pi(\hat{\mathcal A})$ such that $\sum_{j\in J}\hat F_j\in\Pi(\hat{\mathcal A})$,
then there exists a unique functional $p_\alpha(\cdot\mid\hat E):\Pi({\mathcal A})\to [0,1]$ such that
\begin{description}
\item[{\rm (i)}]
$p_\alpha({\bf 1}\mid\hat E)=1$, $p_\alpha(\sum_{j\in J}\alpha(\hat F_j)\mid\hat E)=\sum_{j\in J}p_\alpha(\hat F_j\mid\hat E)$,
\item[{\rm (ii)}]
$p_\alpha(\hat F\mid\hat E)=\frac{\alpha(\hat F)}{\alpha(\hat E)}$
whenever $\hat F\leq \hat E$.
\end{description}}
\noindent
This theorem allows us to single out the unique possible form of the probability $p_\rho(E\&\cdot)$.
\vskip.5pc\noindent
{\bf Theorem 4.2.} {\sl
Let $T$ and $E$ be elementary observables so that $[\hat E,\hat T]={\bf 0}$.
Then $p_\rho(E\&\cdot ):{\mathcal F}(T)\to[0,1]$, $p_\rho(E\& F)=Tr(\rho\hat E\hat F\hat E)$ is the unique functional
which satisfies (C.1.a)-(C.2.a).}
\vskip.5pc\noindent
{\bf Proof.}
If a functional
$p_\rho(E\&\cdot):{\mathcal F}(T)\to [0,1]$ satisfying (C.1.a)-(C.2.a) exists,
then the following functional
$$P_\rho(\cdot\mid\hat E):\hat{\mathcal F}(\hat T)\to [0,1],\quad
P_\rho(\hat F\mid\hat E)=\frac{p_\rho(E\& F)}{p_\rho(E\& {1})}\eqno(4.2)
$$
can be defined.
\par
The set $\hat{\mathcal F}(\hat T)$ is just the set of all projection operators in a
von Neumann algebra as required in theorem 4.1. Namely, the Von Neumann algebra is
$\hat{\mathcal A}(\hat T)=\{\hat T\}'$, and
$\hat{\mathcal F}(\hat T)=\Pi(\hat{\mathcal A}(\hat T))$;
indeed,
given any projection operator $\hat T$, the set
$\hat{\mathcal A}(\hat T)=\{\hat T\}'$ of all bounded linear operators of $\mathcal H$ which commute with
$\hat T$ turns out to be a von Neumann algebra \cite{29}.
Therefore, according to the theory of
von Neumann algebras \cite{29},  $\hat{\mathcal A}(\hat T)$ is the von Neumann algebra generated by $\hat{\mathcal F}(\hat T)$.
\par
 From (C.1.a)-(C.2.a) we deduce that the following statements hold.
\begin{description}
\item[{\rm(i)}]
$P_\rho({\bf 1}\mid \hat E)=p_\rho(E\&1)/p_\rho(E\&1)=1$;
\item[{\rm(ii)}]
making use of (C.2.a) we find
$P_\rho(\sum_{j\in J}\hat F_j\mid\hat E)=\sum_{j\in J}P_\rho(\hat F_j\mid\hat E)$
holds for every countable family $\{\hat F_j\}_{j\in J}$ such that $\sum_{j\in J}\hat F_j\in\hat{\mathcal F}(\hat T)$;
\item[{\rm(iii)}]
if $\hat F\in\hat{\mathcal F}(\hat T)$ and $\hat F\leq\hat E$,
then $P_\rho(\hat F\mid\hat E)=\frac{Tr(\rho\hat F)}{Tr(\rho\hat E)}$ follows from (C.1.a), because
$[\hat F,\hat E]={\bf 0}$ and $\hat F\hat E=\hat F$.
\end{description}
If we put $\alpha(\hat F)=Tr(\rho\hat F)$, we see that the hypotheses of theorem 4.1 holds for this $\alpha$
and $\Pi(\hat{\mathcal A}(\hat T))=\hat{\mathcal F}(\hat T)$. On the other hand
the functional
$P_\rho(\hat F\mid\hat E)=\frac{Tr(\rho\hat E\hat F\hat E)}{Tr(\rho\hat E)}$
satisfies conditions (i) and (ii) of theorem 4.1, therefore it is the unique possibility
for a  functional $P_\rho$ in (4.2), which entails
$p_\rho(E\&F)=P_\rho(\hat F\mid\hat E)p_\rho(E\&{1})=Tr(\rho\hat E\hat F\hat E)$.
\vskip1pc
Theorem 4.2 identifies what is the
unique existing probability $p_\rho(E\& F)$ satisfying (C.1a)-(C.2.a), i.e.,
consistent with the probability prescribed by Quantum Mechanics.
\par
Now, such a unique probability coincides with that determined
by assigning $E$ the value $1$ just when a measurement of the detecting observable $T$ yields the value 1.
Indeed,
if $T$ detects $E$ we have
$P_\rho(\hat F\mid\hat E)=\frac{Tr(\rho\hat E\hat F\hat E)}{Tr(\rho\hat E)}=\frac{Tr(\rho\hat T\hat F\hat E)}{Tr(\rho\hat T)}=
\frac{Tr(\hat T\hat F\hat E\rho)}{Tr(\rho\hat T)}=\frac{Tr(\hat T\hat F\hat T\rho)}{Tr(\rho\hat T)}=
\frac{Tr(\rho\hat F\hat T)}{Tr(\rho\hat T)}$.
\vskip1pc
To accomplish the consistency of assigning $E$ the outcomes of $T$,
requirements (C.1.a)-(C.2.a) must be extended to the case when outcome $0$ is obtained by measuring $T$.
Hence, it must be required that a functional $p_\rho(E'\&\cdot):{\mathcal F}(T)\to[0,1]$ exists such that
\vskip.5pc
\begin{description}
\item[{\rm (C.1.b)}]
if $F\in{\mathcal F}_E(T)\subseteq{\mathcal F}(T)$, i.e. if $[\hat E,\hat F]={\bf 0}$,
then $p_\rho(E'\& F)=Tr(\rho\hat E'\hat F)$.
\item[{\rm (C.2.b)}]
if
$\{\hat F_j\}_{j\in J}\subseteq\hat {\mathcal F}(\hat T)$ is any countable family such that
$\sum_j\hat F_j\equiv \hat F\in\hat{\mathcal F}(\hat T)$, then
$p_\rho(E'\&F)=\sum_{j\in J}p_\rho(E'\& F_j)$;
\end{description}
Repeating the same steps of the proof of theorem 4.2 we can prove the following theorem.
\vskip.5pc\noindent
{\bf Theorem 4.3.} {\sl
Let $T$ and $E$ be elementary observables so that $[\hat E,\hat T]={\bf 0}$.
Then $p_\rho(E'\&\cdot ):{\mathcal F}(T)\to[0,1]$, $p_\rho(E'\& F)=Tr(\rho{\hat E}'\hat F{\hat E}')$ is the unique functional
which satisfies (C.1.b)-(C.2.b).}
\vskip1pc\noindent
Hence,
$P_\rho(\hat F\mid\hat E')=\frac{Tr(\rho\hat E'\hat F\hat E')}{Tr(\rho\hat E')}$
is the unique conditional probability which makes consistent to assign $E$ the outcome of measuring $T$
when such an outcome is $0$.
\vskip.5pc
Yet, we have not finished:
the further condition that the two probabilities $p_\rho(E\&\cdot)$, $p_\rho(E'\&\cdot)$
have to be consistent with each other
and with Quantum Mechanics should be satisfied; i.e., besides (C.1)-(C.2), the following condition should hold.
$$
Tr(\rho\hat F)=p_\rho(E\& F)+p_\rho(E'\& F), \; \forall F\in{\mathcal F}(T).\leqno{\rm (C.3)}
$$
It is immediate to prove that also (C.3) is satisfied.
Indeed,
if $T$ detects $E$ ($\hat T\rho=\hat E\rho$) then $T'=1-T$ detects $E'=1-E$ ($\hat T'\rho=\hat E'\rho$).
So we have that $p_\rho(E\& F)=Tr(\rho \hat T\hat F\hat T)=Tr(\rho\hat F\hat T)$ and
$p_\rho(E'\& F)=Tr(\rho {\hat T}'\hat F{\hat T}')=Tr(\rho\hat F{\hat T}')$
are the only probabilities satisfying (C.1.a)-(C.2.a) and (C.1.b)-(C.2.b) respectively. Therefore
$
p_\rho(E\& F)+p_\rho(E'\& F)=Tr(\rho\hat F\hat T)+ Tr(\rho\hat F{\hat T}')=Tr(\rho\hat F)$.
\vskip1pc
Hence there is a unique occurrence probability for the joint events
``{\sl the value of $E$ is $\eta$}'' and   ``{\sl the value of $F$ is $\phi$}'',
which consistently extends quantum probability to cases where $[\hat E,\hat F]\neq{\bf 0}$.
The role of the detection is just that of yielding the value $\eta$ of $E$ which realizes
the unique consistent probability. Thus,
a language is identified whose sentences are the occurrences of outcomes for observables in ${\mathcal F}(T)$.
The detections provide the language with the value assignment for the sentences
corresponding to the observable $E$, simultaneous to whatever actually measured observable $F$,
also if $[\hat F,\hat E]\neq{\bf 0}$.
The consistency implied by conditions (C.1)-(C.3) ensures the validity, empirical and theoretical,
of the predictions deduced by the language, about their sentences, i.e. about outcomes, statistics and correlations.
\subsection{No Hidden inconsistencies}
Let us suppose that for both the quantum states $\rho_1$ and $\rho_2$ condition (C.3) {\sl does not} hold;
there are
cases where a particular convex combination $\rho=\lambda_1\rho_1+\lambda_2\rho_2$
of these $\rho_1$, $\rho_2$ {\sl does} satisfy (C.3).
In other words, statistical inconsistencies which affect both conditional probabilities $P_{\rho_1}$ and $P_{\rho_2}$
cancel with each other in the convex combination $\lambda_1\rho_1+\lambda_2\rho_2$.
\vskip1pc\noindent
{\bf Example 4.1.}
By making use of two mutually orthonormal vectors $\psi_1,\psi_2\in{\mathcal H}$, we define the two density operators
$\rho_1=\vert\psi_1><\psi_1\vert$, $\rho_2=\vert\psi_2><\psi_2\vert$, and the projection operator
$\hat E=\frac{1}{2}\vert\psi_1+\psi_2><\psi_1+\psi_2\vert$. Then the following relations hold.
$$
E\psi_1=E\psi_2=\frac{1}{2}(\psi_1+\psi_2)\,,\quad \hat E'\psi_1=\frac{1}{2}(\psi_1-\psi_2)\,,
\quad \hat E'\psi_2=\frac{1}{2}(\psi_2-\psi_1)=-\hat E'\psi_1.\eqno(4.3)
$$
If $\rho=\frac{1}{2}(\rho_1+\rho_2)$, by (4.3)
we find
$Tr(\rho\hat E'\hat F\hat E)=\frac{1}{2}(<\psi_1\mid \hat E'\hat F\hat E\psi_1>+<\psi_2\mid \hat E'\hat F\hat E\psi_2>)=
\frac{1}{2}(<\hat E'\psi_1\mid \hat F\hat E\psi_1>+<\hat E'\psi_2\mid \hat F\hat E\psi_2>)=0$, for all $\hat F$.
Similarly, also $Tr(\rho\hat E\hat F\hat E')=0$ holds for all $\hat F$. Then,
$$
Tr(\rho\hat F)=Tr(\rho[\hat E+\hat E']\hat F[\hat E+\hat E'])=Tr(\rho\hat E\hat F\hat E)+Tr(\rho\hat E'\hat F\hat E')
$$
immediately follows.
Hence, (C.3) holds for $\rho=\frac{1}{2}(\rho_1+\rho_2)$ and for all projection operators $\hat F$.
\par\noindent
Then (C.3) must hold also if we fix any $\hat F=\vert(\cos\theta)\psi_1+i(\sin\theta)\psi_2><(cos\theta)\psi_1+i(\sin\theta)\psi_2\vert$,
with $0<\theta<\pi/4$. With this choice of $\hat F$ we find:
\begin{description}
\item[--] $Tr(\rho_1\hat F)=\cos^2\theta$,\quad $Tr(\rho_2\hat F)=\sin^2\theta$;
\item[--]
$Tr(\rho_1\hat E\hat F\hat E)=<\psi_1\mid \hat E\hat F\hat E\psi_1>=\frac{1}{4}<\psi_1+\psi_2\mid\hat F(\psi_1+\psi_2)>=$
\item[] $=\frac{1}{4}(\cos\theta+i\sin\theta)(\cos\theta-i\sin\theta)=\frac{1}{4}$;
similarly, $Tr(\rho_2\hat E\hat F\hat E)=\frac{1}{4}$;
\item[--] similarly
$Tr(\rho_1\hat E'\hat F\hat E')=\frac{1}{4}$, \quad $Tr(\rho_2\hat E'\hat F\hat E')=\frac{1}{4}$
\end{description}
Therefore,
\begin{description}
\item[--]
$Tr(\rho_1\hat E\hat F\hat E)+Tr(\rho_1\hat E'\hat F\hat E')=\frac{1}{2}\neq \cos^2\theta=Tr(\rho_1\hat F)$, and
\item[--]
$Tr(\rho_2\hat E\hat F\hat E)+Tr(\rho_2\hat E'\hat F\hat E')=\frac{1}{2}\neq \sin^2\theta=Tr(\rho_2\hat F)$.
\end{description}
Thus, (C.3) does not hold for both $\rho_1$ and $\rho_2$, but it does hold for $\rho=\frac{1}{2}(\rho_1+\rho_2)$.
\vskip1pc\noindent
The situation shown in example (4.1) would be in contrast with the interpretation of the quantum state $\rho$ in Quantum Mechanics.
A quantum state $\rho$, in Quantum Theory \cite{6},\cite{7},
represents processes which select physical systems.
A selection process is represented by $\rho$ if measurements of every elementary observable $E$
on physical system selected by such a process yield statistics ruled over by the probability
$$
P:{\mathcal E}({\mathcal H})\to[0,1]\,,\quad P(E)=Tr(\rho\hat E).
$$
The set of quantum states is a convex set: if $\lambda_1+\lambda_2=1$ with $\lambda_1,\lambda_2\geq 0$,
then $\rho=\lambda_1\rho_1+\lambda_2\rho_2$ is a density operator, hence a quantum state, if $\rho_1$ and $\rho_2$
are such.
A selection process represented by $\rho=\lambda_1\rho_1+\lambda_2\rho_2$ is not necessarily related to the
selection processes which represent $\rho_1$ and $\rho_2$. However,
a statistical mixture of physical systems selected according to processes corresponding to $\rho_1$ and $\rho_2$,
made up with respective statistical weights $\lambda_1$ and $\lambda_2$, always is represented by the convex combination
$\rho=\lambda_1\rho_1+\lambda_2\rho_2$.
If $\rho_1$, $\rho_2$ and $\rho$ are the quantum states of example 4.1, we have a situation where a consistent value assignment
for $E$ is {\sl impossible} for the selections corresponding to two selections, but it becomes possible by mixing
together the two selections: it is evident that this state of affairs is in contrast with a really consistent value assignment:
{\sl a simple mixture operation should not hide inconsistencies of the component selections}.
\par
Example 4.1 shows that the form of probabilities $p_\rho(E\&F)=Tr(\rho\hat E\hat F\hat E)$ and
$p_\rho(E'\&F)=Tr(\rho\hat E'\hat F\hat E')$
does not prevent from this kind of unacceptable hidden inconsistencies. However,
such a pathology cannot affect the theory developed in the present work:
if the probabilties $p_\rho(E\&F)$ and
$p_\rho(E'\&F)$ are the probabilities ruling over a value assignment by detection, no hidden inconsistencies can occurr,
as the following theorem proves.
\vskip1pc\noindent
{\bf Theorem 4.3.}
{\sl
If $T$ detects $E$ when the system is assigned the state $\rho$, then $T$ detects $E$ when the system is assigned whatever state
$\rho_1$ such that $\rho=\lambda_1\rho_1+\lambda_2\rho_2$, with $\lambda_1>0$.}
\vskip.5pc\noindent
{\bf Proof.}
Let us suppose that $T\quad^\rho\lrcorr E$ where $\rho=\lambda_1\rho_1+\lambda_2\rho_2$ with $\lambda_1>0$. Then Prop. 2.1
implies that
$$
[\hat T,\hat E]={\bf 0},\;\hbox{hence }\hat E\hat T\leq\hat T,\;
\hat E\hat T\leq E\,,\quad \hat E\hat T\rho =\hat T\rho=\hat E\rho\,.\eqno(4.3)
$$
By (4.3) we have $Tr(\rho[\hat T-\hat E\hat T])=0$, i.e.
$Tr([\lambda_1\rho_1+\lambda_2\rho_2][\hat T-\hat E\hat T])=0$, which implies
$\lambda_1 Tr(\rho_1[\hat T-\hat E\hat T]) +\lambda_2Tr(\rho_2[\hat T-\hat E\hat T])=0$;
thereby, $Tr(\rho_1[\hat T-\hat E\hat T])=0$ follows, because $\hat T-\hat E\hat T\geq{\bf 0}$.
Therefore, $[\hat T-\hat E\hat T]\rho_1=0$ holds.
\par\noindent
Similarly we can deduce $[\hat E-\hat E\hat T]\rho_1=0$; thus $\hat T\rho_1=\hat E\rho_1$.
\vskip1pc\noindent
In other words,
consistency is not lost by any possible refinement of the selection corresponding to the quantum state $\rho$.


\end{document}